\documentclass[conference]{IEEEtran}
\IEEEoverridecommandlockouts

\usepackage{cite}
\usepackage{amsmath,amssymb,amsfonts}
\usepackage{algorithmic}
\usepackage{graphicx}
\usepackage{textcomp}
\usepackage{xcolor}
\usepackage{colortbl}
\usepackage{pifont} 
\newcommand{\cmark}{\ding{51}} 

\usepackage{booktabs}
\usepackage{multirow}
\usepackage[compatibility=false]{caption}
\usepackage{amsmath} 
\newcommand{\std}[2]{#1\textsubscript{\scriptsize\,$\pm$\,#2}}

\usepackage{subfig}
\definecolor{lightgrayrow}{gray}{0.92}
\def\BibTeX{{\rm B\kern-.05em{\sc i\kern-.025em b}\kern-.08em
    T\kern-.1667em\lower.7ex\hbox{E}\kern-.125emX}}
\begin{document}

\title{Clinically Informed Acoustic-Aware Attribute Modeling for Respiratory Sound Classification
\thanks{Hubei Province Technological Innovation Major Project (No. 2022BCA041).}
}

\author{
\IEEEauthorblockN{
Xinhong Li\textsuperscript{1}, 
Li Xiao\textsuperscript{1}, 
Ziyang Chen\textsuperscript{1}, 
Yuhong Yang\textsuperscript{1}, 
Weiping Tu\textsuperscript{1,2,*}
\thanks{*Corresponding author: Weiping Tu (tuweiping@whu.edu.cn)}
}
\IEEEauthorblockA{\textsuperscript{1}\textit{School of Computer Science, Wuhan University, Wuhan, China}}
\IEEEauthorblockA{\textsuperscript{2}\textit{Hubei Key Laboratory of Multimedia and Network Communication Engineering, Wuhan University, Wuhan, China}}
}

\maketitle

\maketitle

\maketitle

\begin{abstract}
Automated respiratory sound classification (RSC) aims to identify normal and adventitious respiratory sounds, such as crackles and wheezes. Most existing methods formulate RSC as a flat four-class classification problem over \textit{normal}, \textit{crackle}, \textit{wheeze}, and \textit{both}. However, unlike conventional classification tasks with semantically independent and mutually exclusive categories, the commonly used four-class RSC labels have a compositional structure and can be represented by two clinically meaningful sound attributes: crackle presence and wheeze presence. Existing flat classifiers do not explicitly exploit this structure, which may increase the learning burden by treating the four labels as independent targets. Meanwhile, crackles and wheezes exhibit distinct temporal characteristics, with sparse transient crackle events and sustained continuous wheeze patterns, making a single temporal aggregation strategy less suitable for modeling different adventitious sound evidence.
To address these limitations, we propose an Acoustic-Aware Asymmetric Framework (AAAF) for RSC. AAAF reformulates four-class recognition as two sound-attribute detection tasks and recovers the final category through logical composition during inference. Built upon a pretrained BEATs encoder, AAAF employs branch-specific multi-layer fusion and asymmetric temporal aggregation to capture localized crackle evidence and sustained wheeze patterns. Auxiliary normal--abnormal supervision and a batch-adaptive dynamic loss are further introduced to improve global discrimination and robustness under imbalanced labels. Experiments on ICBHI and SPRSound demonstrate that AAAF improves overall RSC performance and generalizes across datasets, achieving a state-of-the-art four-class Score of 65.30\% on ICBHI.

\end{abstract}

\begin{IEEEkeywords}
Respiratory sound classification, computer-aided auscultation, adventitious respiratory sounds, task reformulation, acoustic-aware temporal aggregation.
\end{IEEEkeywords}

\section{Introduction}
Respiratory diseases pose a significant global health burden, motivating the need for accessible and objective early screening methods. Pulmonary auscultation provides a non-invasive way to assess respiratory conditions by listening to respiratory sounds, including normal breath sounds and adventitious sounds such as crackles and wheezes. Crackles are discontinuous explosive sounds associated with airway secretions or reopening of collapsed airways~\cite{piirila1995crackles}, while wheezes are continuous musical sounds related to airway obstruction~\cite{meslier1995wheezes}. However, reliable auscultation requires trained clinicians and may be difficult to provide consistently in telemedicine, community screening, and resource-limited settings~\cite{dong25e_interspeech}. These limitations have motivated increasing interest in automated respiratory sound classification (RSC).

\begin{figure}[t]
  \centering
  \includegraphics[width=0.45\textwidth]{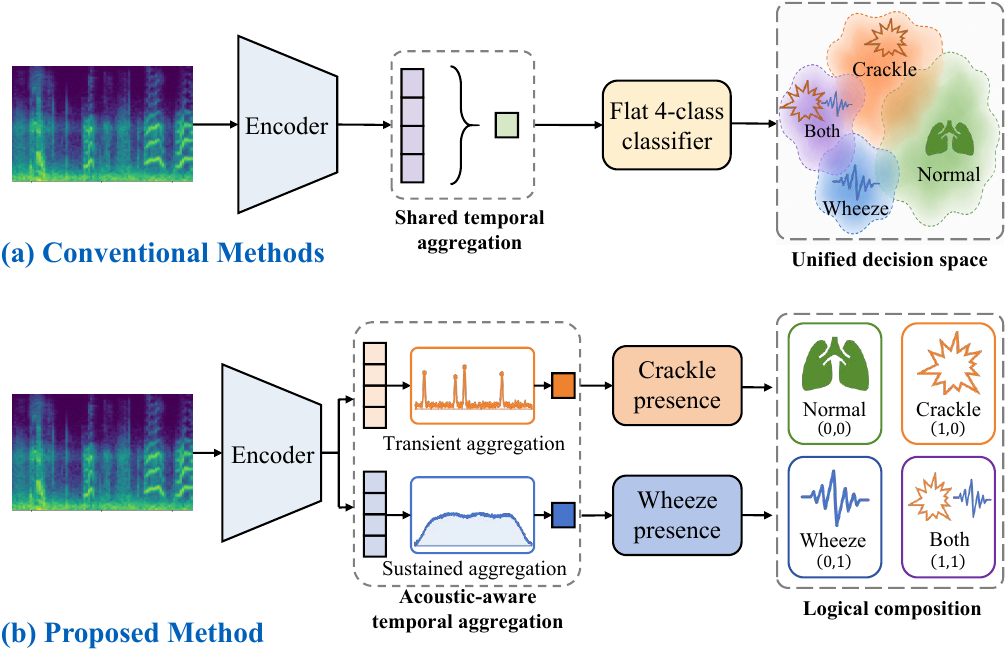}
  \caption{Motivation of AAAF: flat four-class RSC versus crackle--wheeze attribute modeling with acoustic-aware temporal aggregation.}
  \label{fig:intro}
\end{figure}

In this line of research, most existing deep learning methods formulate RSC as a flat four-class classification problem, assigning each respiratory event to one of four categories: \textit{normal}, \textit{crackle}, \textit{wheeze}, and \textit{both}~\cite{moummad2023pretraining,xiao24_interspeech,kim2024stethoscope,jeong25_interspeech}, as illustrated in Fig.~\ref{fig:intro}(a). Although this formulation is simple and widely adopted, it does not fully reflect the label semantics of respiratory sound benchmarks. Unlike conventional classification tasks whose categories usually correspond to mutually exclusive semantic concepts, these four labels can be represented by two adventitious sound presence attributes: crackle presence and wheeze presence. Specifically, \textit{normal} indicates the absence of both attributes, while \textit{both} indicates their joint presence within the same respiratory cycle. Therefore, treating the four labels as independent targets imposes an artificial mutual-exclusion assumption and requires the model to learn four separate decision boundaries, rather than explicitly sharing crackle- and wheeze-related evidence across categories. This may increase the learning burden and obscure the underlying compositional structure of RSC labels.

Beyond this task-formulation limitation, effective RSC models should also account for the distinct temporal patterns of crackles and wheezes. Crackles are short, discontinuous, and transient events that usually appear within localized time regions~\cite{doi:10.1056/NEJMra1302901}, whereas wheezes are sustained harmonic sounds that extend over longer temporal intervals~\cite{pasterkamp1997respiratory}. Although recent CNN-based~\cite{9630091,9669787,9746182,9747014} and Transformer-based~\cite{10782363,bae2023patch,wei25_interspeech,10888016} methods have improved RSC with stronger neural encoders and representation learning strategies, most of them still aggregate acoustic evidence through a shared temporal modeling pipeline, as illustrated in Fig.~\ref{fig:intro}(a). Such a common aggregation strategy tends to produce a single event representation, making it difficult to emphasize sparse transient crackle cues while also preserving sustained wheeze patterns. These observations motivate acoustic-aware temporal modeling with attribute-specific asymmetric aggregation.

To address these limitations, we propose an Acoustic-Aware Asymmetric Framework (AAAF) for respiratory sound classification, as shown in Fig.~\ref{fig:intro}(b). AAAF reformulates flat four-class recognition as two adventitious sound presence detection tasks, i.e., crackle presence and wheeze presence, and recovers the final category through logical composition during inference. Built upon a pretrained BEATs encoder, AAAF introduces two attribute-specific branches with learnable multi-layer fusion. The crackle branch emphasizes localized transient responses, while the wheeze branch captures sustained temporal patterns. Considering the severe label imbalance in respiratory sound datasets, auxiliary normal--abnormal supervision and a batch-adaptive dynamic loss are further incorporated to improve global discrimination and robust optimization.

The main contributions of this work are summarized as follows:
\begin{itemize}
    \item  We revisit RSC from a compositional label-structure perspective and reformulate flat four-class recognition as two adventitious sound presence detection tasks, enabling logical composition of the final prediction during inference.
    
    \item We propose AAAF, an acoustic-aware asymmetric framework built upon hierarchical BEATs representations, with branch-specific multi-layer fusion and attribute-specific temporal aggregation for crackle and wheeze detection.
    
    \item We introduce auxiliary normal--abnormal supervision and a batch-adaptive dynamic loss to improve global discrimination and robustness under imbalanced RSC labels. Extensive experiments on ICBHI and SPRSound demonstrate the effectiveness and cross-dataset generalization of AAAF.
\end{itemize}

\begin{figure*}[t]
  \centering
  \includegraphics[width=0.95\textwidth]{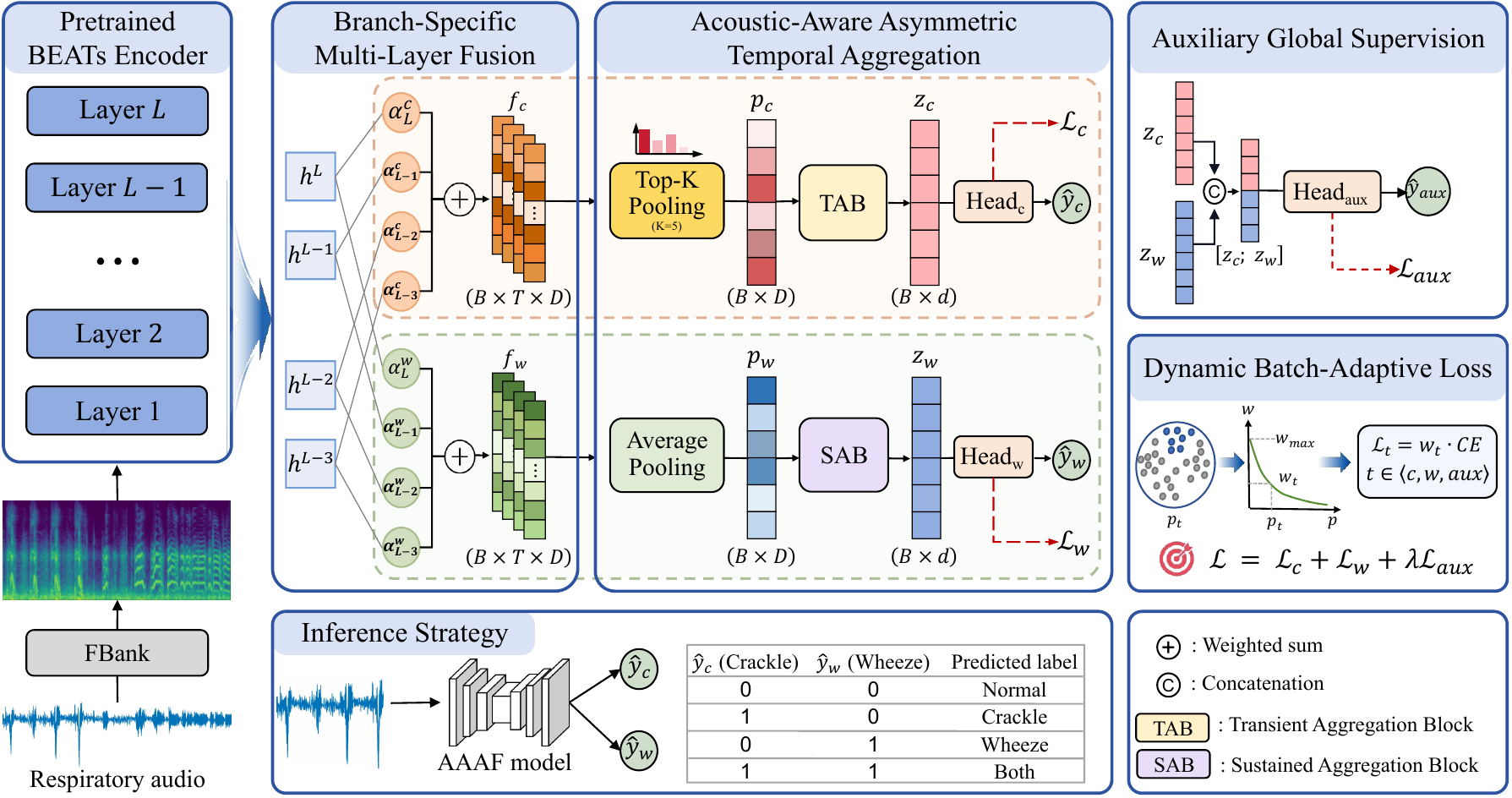}
  \caption{Overview of the proposed AAAF. Hierarchical BEATs representations are fused and fed into two attribute-specific branches for crackle and wheeze presence detection. Temporal Top-$K$ pooling emphasizes localized crackle evidence, while average pooling with channel attention captures sustained wheeze patterns. The final four-class prediction is obtained by logical composition, with auxiliary normal--abnormal supervision and batch-adaptive dynamic loss for joint optimization.}
  \label{fig:model}
\end{figure*}

\section{Method}

As illustrated in Fig.~\ref{fig:model}, the proposed Acoustic-Aware Asymmetric Framework (AAAF) reformulates flat four-class respiratory sound classification (RSC) as crackle- and wheeze-presence detection, and recovers the final category through logical composition. Built upon hierarchical BEATs representations, AAAF consists of five components: task reformulation and inference composition, branch-specific multi-layer fusion, acoustic-aware asymmetric temporal aggregation, auxiliary global supervision, and dynamic batch-adaptive loss.

\subsection{Task Reformulation and Inference Composition}

Given a respiratory event $x$, conventional RSC assigns it to a single label $y \in \{0,1,2,3\}$, corresponding to \textit{normal}, \textit{crackle}, \textit{wheeze}, and \textit{both}. To exploit the compositional structure of this label space, we represent each category using two binary sound attributes: crackle presence and wheeze presence. Specifically,

\begin{equation}
y_c = \mathbb{I}(y \in \{1,3\}), \qquad
y_w = \mathbb{I}(y \in \{2,3\}),
\end{equation}
where $\mathbb{I}(\cdot)$ denotes the indicator function, which returns 1 if the condition is true and 0 otherwise. Here, $y_c$ and $y_w$ indicate whether crackle and wheeze are present in the respiratory event. Thus, normal, crackle, wheeze, and both correspond to $(0,0)$, $(1,0)$, $(0,1)$, $(1,1)$, respectively.

AAAF estimates the posterior probabilities $\hat{p}_c=P(y_c=1|x)$ and $\hat{p}_w=P(y_w=1|x)$ using two attribute-specific branches. During inference, binary predictions are obtained by applying thresholds $\tau_c$ and $\tau_w$:

\begin{equation}
\hat{y}_c = \mathbb{I}(\hat{p}_c > \tau_c), \qquad
\hat{y}_w = \mathbb{I}(\hat{p}_w > \tau_w).
\end{equation}

The final four-class prediction is recovered by logical composition of $\hat{y}_c$ and $\hat{y}_w$:
\begin{equation}
\hat{y} =
\begin{cases}
0, & \hat{y}_c=0, \hat{y}_w=0,\\
1, & \hat{y}_c=1, \hat{y}_w=0,\\
2, & \hat{y}_c=0, \hat{y}_w=1,\\
3, & \hat{y}_c=1, \hat{y}_w=1.
\end{cases}
\end{equation}

\subsection{Acoustic-Aware Asymmetric Architecture}

\subsubsection{Branch-Specific Multi-Layer Fusion}
Pretrained audio encoders provide hierarchical representations, where different layers may encode complementary acoustic information. Lower and intermediate layers tend to preserve fine-grained spectro-temporal details, while deeper layers capture more abstract event-level patterns. Since crackles and wheezes exhibit different temporal characteristics, directly using only the final-layer representation may be suboptimal for both attributes.

Given an input respiratory event $x$, we feed it into a pretrained BEATs encoder and extract frame-level hidden states from selected Transformer layers. Let $\mathbf{h}^l \in \mathbb{R}^{T \times D}$ denote the hidden representation from the $l$-th layer, where $T$ is the number of frames and $D$ is the feature dimension. For each branch $b \in {c,w}$, the selected hidden states are fused as
\begin{equation}
\mathbf{f}_b = \sum_{l \in S_b} \alpha_b^l \mathbf{h}^l.
\end{equation}
where $S_b$ denotes the selected layer set for branch $b$, and $\alpha_b^l$ is a learnable layer-fusion weight. 
In our implementation, both branches use the last four BEATs layers as candidates, i.e., $S_c=S_w=\{L-3,L-2,L-1,L\}$, while maintaining independent fusion weights. This design allows the crackle and wheeze branches to adaptively emphasize different hierarchical representations before entering their corresponding acoustic-aware temporal aggregation branches.

\subsubsection{Acoustic-Aware Asymmetric Temporal Aggregation}

After branch-specific multi-layer fusion, AAAF processes the crackle and wheeze representations with two attribute-specific branches. Although both branches are derived from the same input event, they adopt different temporal aggregation and feature transformation strategies according to the acoustic characteristics of their target attributes.

\textbf{Crackle branch.}
Crackles are sparse and transient events whose discriminative evidence is often concentrated in short localized regions. Direct average pooling over all frames may dilute these transient responses with surrounding breath sounds. Therefore, for the fused crackle feature $\mathbf{f}_c \in \mathbb{R}^{T \times D}$, we apply temporal Top-$K$ pooling to preserve the strongest local responses:
\begin{equation}
\mathbf{u}_c = \mathrm{TopKPool}(\mathbf{f}_c, K),
\end{equation}
where $\mathrm{TopKPool}(\cdot)$ selects the top $K$ temporal responses and averages them into a fixed-dimensional representation. This allows the crackle branch to focus on the most salient transient evidence rather than the entire respiratory event. The pooled feature $\mathbf{u}_c$ is then fed into a Transient Aggregation Block (TAB), implemented as an MLP projection, to obtain the crackle-oriented representation:
\begin{equation}
\mathbf{z}_c = \mathrm{TAB}(\mathbf{u}_c),
\end{equation}
A binary classification head is finally applied to predict crackle presence:
\begin{equation}
\hat{p}_c = \sigma(\mathrm{Head}_c(\mathbf{z}_c)).
\end{equation}

\textbf{Wheeze branch.}
Wheezes are sustained and continuous sounds that extend over longer temporal intervals. To capture such long-duration abnormal patterns, we apply temporal average pooling to the fused wheeze feature $\mathbf{f}_w \in \mathbb{R}^{T \times D}$:
\begin{equation}
\mathbf{u} = \frac{1}{T}\sum_{t=1}^{T} \mathbf{f}_w(t),
\end{equation}
The pooled feature $u_w$ is processed by a Sustained Aggregation Block (SAB), which consists of an MLP followed by channel attention to recalibrate feature channels for wheeze detection:
\begin{equation}
\mathbf{z}_w = \mathrm{SAB}(\mathbf{u}_w),
\end{equation}
A separate binary classification head predicts wheeze presence:
\begin{equation}
\hat{p}_w = \sigma(\mathrm{Head}_w(\mathbf{z}_w)).
\end{equation}

This asymmetric temporal aggregation allows the crackle branch to preserve localized transient evidence, while enabling the wheeze branch to capture sustained temporal patterns. The resulting representations $\mathbf{z}_c$ and $\mathbf{z}_w$ are used for attribute-specific predictions and further combined for auxiliary global supervision.

\subsection{Global Supervision and Adaptive Optimization}

\subsubsection{Auxiliary Global Supervision}
Although the two attribute-specific branches focus on crackle and wheeze presence separately, reliable RSC also requires global discrimination between normal respiratory sounds and abnormal acoustic events. To provide this global constraint, we introduce an auxiliary normal--abnormal supervision task.

The auxiliary target is defined as
\begin{equation}
y_a = \mathbb{I}(y \neq 0),
\end{equation}
where $y=0$ denotes the normal class. The branch-specific representations $\mathbf{z}_c$ and $\mathbf{z}_w$ are concatenated and fed into an auxiliary binary classifier:
\begin{equation}
\hat{p}_a = \sigma(\mathrm{Head}_a(\mathbf{z}_c \oplus \mathbf{z}_w).
\end{equation}
where $\oplus$ denotes feature concatenation. This auxiliary objective encourages the two attribute-specific branches to preserve global normal--abnormal discriminability while learning crackle- and wheeze-oriented representations.

\subsubsection{Dynamic Batch-Adaptive Loss}
Respiratory sound datasets are often highly imbalanced, with abnormal attributes and co-occurring events being under-represented. Since the positive ratio of each binary task may vary across mini-batches, fixed class weights may not adequately reflect the current training distribution. We therefore adopt a batch-adaptive weighting strategy for the crackle, wheeze, and auxiliary normal--abnormal tasks. For each task $t \in \{c,w,a\}$, let $r_t$ denote the positive ratio in the current mini-batch. The positive-class weight is computed as
\begin{equation}
\omega_t^{+} =
\min
\left(
\omega_{\max}^{t},
\max
\left(
1,
\frac{1-r_t}{r_t+\epsilon}
\right)
\right),
\end{equation}

where $\epsilon$ is a small constant for numerical stability and $\omega_{\max}^{t}$ prevents excessively large weights under extreme imbalance. The negative-class weight is fixed to $\omega_t^{-}=1$.

For each binary task, the weighted cross-entropy loss is defined as

\begin{equation}
\mathcal{L}^{t}_{\mathrm{WCE}}
=
-\frac{1}{B}
\sum_{i=1}^{B}
\left[
\omega_t^{+} y_{t,i} \log p_{t,i}
+
(1-y_{t,i}) \log (1-p_{t,i})
\right],
\end{equation}
where $B$ is the batch size, $y_{t,i}$ is the binary target, and $p_{t,i}$ is the predicted positive probability for sample $i$.

The final training objective is
\begin{equation}
\mathcal{L}
=
\mathcal{L}^{c}_{\mathrm{WCE}}
+
\mathcal{L}^{w}_{\mathrm{WCE}}
+
\lambda
\mathcal{L}^{a}_{\mathrm{WCE}}.
\end{equation}
where $\lambda$ controls the contribution of the auxiliary normal--abnormal supervision.

\section{Experimental Setup}

\begin{table*}[t!]
\centering
\small
\setlength{\tabcolsep}{5.2pt}
\renewcommand{\arraystretch}{1.15}
\caption{Performance comparison on the ICBHI 2017 benchmark. \textbf{Bold} and \underline{underline} indicate the best and second-best results, respectively. IN, AS, and LA denote ImageNet~\cite{5206848}, AudioSet~\cite{7952261}, and LAION-Audio-630K~\cite{10095969} pretraining.}
\label{tab:icbhi_comparison}

\begin{tabular}{c l c c c c c c}
\toprule
& Method & Architecture & Pretrain & Venue & $S_p$ (\%) & $S_e$ (\%) & Score (\%) \\
\midrule

\multirow{10}{*}{\rotatebox{90}{\textbf{4-class eval.}}}
& Patch-Mix~\cite{bae2023patch}  & AST & IN + AS & \textit{INTERSPEECH'23} & 81.66 & 43.07 & 62.37 \\
& SG-SCL~\cite{kim2024stethoscope} & AST & IN + AS & \textit{ICASSP'24} & 79.87 & 43.55 & 61.71 \\
& RepAugment~\cite{10782363} & AST & IN + AS & \textit{EMBC'24} & 82.47 & 40.55 & 61.51 \\
& LungAdapter~\cite{xiao24_interspeech} & AST & IN + AS & \textit{INTERSPEECH'24} & 80.43 & 44.37 & 62.40 \\
& BTS~\cite{kim24f_interspeech} & CLAP & LA & \textit{INTERSPEECH'24} & 81.40 & 45.67 & 63.54 \\
& AMG-SCL~\cite{77b52a84c80c4423bb6d9cedf8ea2963} & AST & IN + AS & \textit{JBHI'25} & 76.18 & \underline{47.91} & 62.04 \\
& BTS-d~\cite{toikkanen25_interspeech} & CLAP & LA & \textit{INTERSPEECH'25} & \underline{82.89} & 45.90 & 64.39 \\
& PAFA~\cite{jeong25_interspeech} & BEATs & AS & \textit{INTERSPEECH'25} & 82.50 & 47.63 & \underline{64.84} \\
& BTS-CARD~\cite{koo2026empowering} & CLAP & LA & \textit{ICASSP'26} & \textbf{84.42} & 44.83 & 64.63 \\
\cmidrule(lr){2-8}
& \textbf{AAAF [ours]} & BEATs & AS & -- & \std{80.25}{3.57} & \std{\textbf{50.35}}{3.15} & \std{\textbf{65.30}}{0.50} \\

\midrule

\multirow{6}{*}{\rotatebox{90}{\textbf{2-class eval.}}}
& CNN-MoE~\cite{9372748} & C-DNN & -- & \textit{JBHI'21} & 72.40 & 37.50 & 54.10 \\
& Nguyen \textit{et al.}~\cite{9729496} & ResNet50 & IN & \textit{TBME'22} & 79.34 & 50.14 & 64.74 \\
& Patch-Mix~\cite{bae2023patch} & AST & IN + AS & \textit{INTERSPEECH'23} & \textbf{81.66} & 55.77 & 68.71 \\
& SG-SCL~\cite{kim2024stethoscope} & AST & IN + AS & \textit{ICASSP'24} & \underline{79.87} & 57.97 & 68.93 \\
& PAFA~\cite{jeong25_interspeech} & BEATs & AS & \textit{INTERSPEECH'25} & 74.87 & \underline{68.29} & \textbf{72.08} \\
\cmidrule(lr){2-8}
& \textbf{AAAF [ours]} & BEATs & AS & -- & \std{74.01}{4.07} & \std{\textbf{68.89}}{2.90} & \std{\underline{71.45}}{0.87} \\
\bottomrule
\end{tabular}
\end{table*}

\begin{table*}[t!]
\centering
\small
\setlength{\tabcolsep}{4.8pt}
\renewcommand{\arraystretch}{1.15}
\caption{External validation results on the SPRSound dataset. $\dagger$ denotes results reproduced by us under the same evaluation protocol. Pretraining abbreviations follow Table~\ref{tab:icbhi_comparison}.}
\label{tab:sprsound_comparison}

\begin{tabular}{l c c ccc ccc}
\toprule
\multirow{2}{*}{Method} 
& \multirow{2}{*}{Architecture} 
& \multirow{2}{*}{Pretrain} 
& \multicolumn{3}{c}{4-class eval.} 
& \multicolumn{3}{c}{2-class eval.} \\
\cmidrule(lr){4-6} \cmidrule(lr){7-9}
& & & $S_p$ (\%) & $S_e$ (\%) & Score (\%) 
& $S_p$ (\%) & $S_e$ (\%) & Score (\%) \\
\midrule

ResNet-50\textsuperscript{\textdagger} 
& CNN & IN 
& 91.33 & 82.81 & 87.07
& 89.72 & 90.07 & 89.90 \\

EfficientNet-B0\textsuperscript{\textdagger} 
& CNN & IN 
& \textbf{91.68} & 83.57 & 87.63
& 90.30 & 91.15 & 90.72 \\

AST\textsuperscript{\textdagger} 
& AST & IN + AS 
& 90.85 & 84.65 & 87.75
& 89.80 & 92.57 & 91.19 \\

BEATs\textsuperscript{\textdagger} 
& BEATs & AS 
& \underline{91.39} & \underline{85.36} & \underline{88.38}
& 90.06 & \textbf{93.99} & 92.02 \\

SCL~\cite{moummad2023pretraining} 
& CNN & AS 
& 90.49 & 80.69 & 85.59
& -- & -- & -- \\

CNN-TSA~\cite{fraihi2025improving} 
& CNN & AS 
& 91.38 & 76.03 & 83.70
& -- & -- & -- \\

DPAST-LS~\cite{zhang2026dpast} 
& AST & IN + AS 
& -- & -- & --
& \textbf{93.98} & 92.06 & \textbf{93.02} \\

\midrule

\textbf{AAAF [ours]} 
& BEATs & AS 
& \std{90.95}{0.84} & \std{\textbf{86.13}}{0.89} & \std{\textbf{88.54}}{0.38}
& \std{90.95}{1.13} & \std{\underline{93.47}}{1.17} & \std{\underline{92.21}}{0.14} \\

\bottomrule
\end{tabular}
\end{table*}

\subsection{Datasets}

\textbf{ICBHI.}
The ICBHI 2017 dataset contains 920 recordings from 126 subjects and 6,898 annotated respiratory cycles~\cite{rocha2017alpha}. Following the official subject-wise split, we use 4,142 cycles for training and 2,756 for testing, with four classes: Normal, Crackle, Wheeze, and Both.

\textbf{SPRSound.}
SPRSound contains 2,683 recordings and 9,089 respiratory sound events from 292 participants~\cite{9880516}. We use its official split, with 6,656 training and 2,433 testing events, for external validation. Its seven event-level categories are mapped to the same four-class space: Coarse/Fine Crackle to Crackle, Rhonchi/Wheeze/Stridor to Wheeze, and Wheeze+Crackle to Both, while Normal remains unchanged.

\subsection{Implementation Details}

All audio recordings are resampled to 16 kHz and cropped or padded to 5-second clips before being fed into the pretrained BEATs encoder. The BEATs backbone is initialized with AudioSet-pretrained weights. We optimize AAAF using Adam with a batch size of 32, weight decay of $1\times10^{-6}$, and cosine learning-rate scheduling. Differential learning rates are used for different modules: $5\times10^{-5}$ for the BEATs encoder, $1\times10^{-3}$ for the classification branches, and $2\times10^{-3}$ for the layer-fusion parameters. The auxiliary loss weight $\lambda$ is set to 0.2. TThe branch-specific thresholds $\tau_c$ and $\tau_w$ are selected via grid search without using the test labels and are kept fixed across all random seeds. All experiments are repeated with five random seeds, and results are reported as mean $\pm$ standard deviation.

\subsection{Evaluation Metrics}

Following the standard ICBHI protocol, we report Specificity (\(S_p\)), Sensitivity (\(S_e\)), and \(\mathrm{Score}=(S_p+S_e)/2\). For four-class evaluation, Normal is treated as the negative class and the three adventitious classes as positive. For binary evaluation, the three adventitious classes are merged into Abnormal.

\section{Results}

\subsection{Main Comparison}

Table~\ref{tab:icbhi_comparison} compares AAAF with recent state-of-the-art methods on the ICBHI benchmark. In the four-class evaluation, AAAF achieves the best overall Score of 65.30\% and the highest Sensitivity ($S_e$) of 50.35\%, while maintaining competitive Specificity ($S_p$). This demonstrates the effectiveness of attribute-based reformulation and acoustic-aware asymmetric modeling in recognizing adventitious respiratory sounds. For the two-class Normal-vs-Abnormal evaluation, AAAF achieves a Score of 71.45\% and the highest Sensitivity of 68.89\%, demonstrating strong abnormal sound detection capability.

To evaluate cross-dataset generalization, we further validate AAAF on SPRSound, as shown in Table~\ref{tab:sprsound_comparison}. Compared with the reproduced BEATs baseline, AAAF improves the four-class and two-class Scores from 88.38\% to 88.54\% and from 92.02\% to 92.21\%, respectively. These results demonstrate that AAAF remains effective under a different dataset and event taxonomy.

\subsection{Ablation and Component Analysis}

\subsubsection{Effectiveness of Proposed Components}
Table~\ref{tab:ablation_components} reports the ablation results of AAAF on the ICBHI four-class evaluation. The vanilla BEATs baseline, which directly uses the final-layer representation for four-class classification without the proposed components, achieves a Score of 63.49\%.
Adding branch-specific multi-layer fusion improves Sensitivity from 48.21\% to 50.55\%, although it slightly decreases Specificity. Adding acoustic-aware asymmetric temporal aggregation (AATA) further improves the Score to 64.29\%, indicating the benefit of using different aggregation strategies for transient crackles and sustained wheezes.

The auxiliary normal--abnormal supervision further improves the Score to 64.49\%, while the dynamic batch-adaptive loss mainly enhances Sensitivity, reaching 51.78\%. Combining all components yields the best Score of 65.30\%, with balanced Specificity and Sensitivity. These results suggest that the proposed components are complementary: multi-layer fusion enriches BEATs representations, asymmetric temporal aggregation adapts to different crackle and wheeze patterns, auxiliary supervision preserves global abnormality discrimination, and dynamic loss improves robustness under imbalanced labels.

\begin{table}[t!]
\centering
\footnotesize 
\setlength{\tabcolsep}{2.5pt} 
\renewcommand{\arraystretch}{1.1} 
\caption{Ablation study of AAAF components. Multi: multi-layer fusion; AATA: acoustic-aware asymmetric temporal aggregation; Aux: auxiliary task; DyLoss: dynamic batch-adaptive loss.}
\label{tab:ablation_components}
\begin{tabular}{@{} cccc ccc @{}}
\toprule
\multicolumn{4}{c}{Modules} & \multicolumn{3}{c}{Performance (\%)} \\
\cmidrule(lr){1-4}\cmidrule(lr){5-7}
Multi & AATA & Aux & DyLoss & $S_p$ & $S_e$ & Score \\
\midrule

 &  &  &  &
\std{78.77}{3.07} & \std{48.21}{2.32} & \std{63.49}{1.08} \\
\cmark &  &  &  &
\std{76.35}{2.60} & \std{\underline{50.55}}{2.24} & \std{63.45}{0.77} \\

\cmark & \cmark &  &  &
\std{\textbf{82.84}}{3.03} & \std{45.74}{3.30} & \std{64.29}{0.58} \\

\cmark & \cmark & \cmark &  &
\std{\underline{80.59}}{3.60} & \std{48.39}{3.47} & \std{64.49}{0.08} \\

\cmark & \cmark &  & \cmark &
\std{77.48}{0.85} & \std{\textbf{51.78}}{1.95} & \std{\underline{64.63}}{0.83} \\

\cmidrule(lr){1-7} 
\cmark & \cmark & \cmark & \cmark &
\std{80.25}{3.57} & \std{50.35}{3.15} & \std{\textbf{65.30}}{0.50} \\

\bottomrule
\end{tabular}
\end{table}

\subsubsection{Impact of Layer Selection}
\begin{table}[t]
    \centering
    \caption{Comparison of BEATs layer selection strategies for hierarchical feature extraction.}
    \label{tab:layer_selection}
    \resizebox{\columnwidth}{!}{%
    \begin{tabular}{l c c c}
        \toprule
        Layer Selection & $S_p$ (\%) & $S_e$ (\%) & Score (\%) \\
        \midrule
        Last Layer      & \std{76.81}{2.18} & \std{\textbf{51.35}}{3.20} & \std{64.08}{0.81} \\
        Interval Layers & \std{\underline{79.59}}{3.13} & \std{49.21}{3.67} & \std{\underline{64.40}}{0.67} \\
        All Layers      & \std{77.39}{2.23} & \std{49.21}{2.65} & \std{63.30}{0.62} \\
        Ours            & \std{\textbf{80.25}}{3.57} & \std{\underline{50.35}}{3.15} & \std{\textbf{65.30}}{0.50} \\
        \bottomrule
    \end{tabular}
    } 
\end{table}
Table~\ref{tab:layer_selection} compares different BEATs layer selection strategies for branch-specific feature fusion. Using only the final layer obtains a Score of 64.08\%, showing that the final representation alone may not fully capture the fine-grained acoustic cues needed for RSC. Selecting interval layers (layers 3, 6, 9, and 12) improves the Score to 64.40\%, suggesting that multi-level representations provide complementary information. 

However, aggregating all layers does not further improve performance and yields a lower Score of 63.30\%, likely because early or less relevant layers introduce noisy information. The proposed strategy, which fuses the last four BEATs layers, achieves the best Score of 65.30\%. This indicates that deeper hierarchical representations provide a favorable balance between representation diversity and feature consistency.

\subsubsection{Impact of Top-$K$ Pooling}

\begin{table}[t]
\centering
\footnotesize
\setlength{\tabcolsep}{6pt}
\renewcommand{\arraystretch}{1.12}
\caption{Impact of Top-$K$ pooling in the crackle branch.}
\label{tab:topk}
\begin{tabular}{c c c c}
\toprule
$K$ & $S_p$ (\%) & $S_e$ (\%) & Score (\%) \\
\midrule
1  & \std{82.13}{4.03} & \std{46.29}{3.77} & \std{64.21}{0.84} \\
3  & \std{79.38}{2.68} & \std{50.09}{1.74} & \std{64.74}{0.64} \\
\rowcolor{lightgrayrow}
5  & \std{80.25}{3.57} & \std{\textbf{50.35}}{3.15} & \std{\textbf{65.30}}{0.50} \\
7  & \std{79.54}{7.57} & \std{48.84}{6.35} & \std{64.19}{1.00} \\
9  & \std{\textbf{84.23}}{6.32} & \std{45.18}{7.19} & \std{64.70}{0.94} \\
11 & \std{80.11}{5.15} & \std{49.02}{4.65} & \std{64.57}{0.50} \\
13 & \std{79.73}{3.18} & \std{48.89}{2.97} & \std{64.31}{0.45} \\
\bottomrule
\end{tabular}
\end{table}

Table~\ref{tab:topk} investigates the effect of the Top-$K$ pooling parameter in the crackle branch. Small values of $K$ focus on only the most prominent frames, which may miss useful contextual cues, while overly large values may introduce irrelevant background information and weaken transient crackle responses. The results show that a moderate value provides a better balance between sensitivity and specificity.

Among the tested settings, $K=5$ achieves the best overall Score of 65.30\%, indicating that selecting a limited number of salient temporal frames is effective for preserving transient crackle evidence. When $K$ is further increased, the performance does not continue to improve, suggesting that excessive temporal aggregation may dilute localized crackle-related cues. Therefore, we set $K=5$ in all experiments.

\subsubsection{Frame-wise Branch Responses}

\begin{figure}[t]
    \centering
    \includegraphics[width=0.95\linewidth]{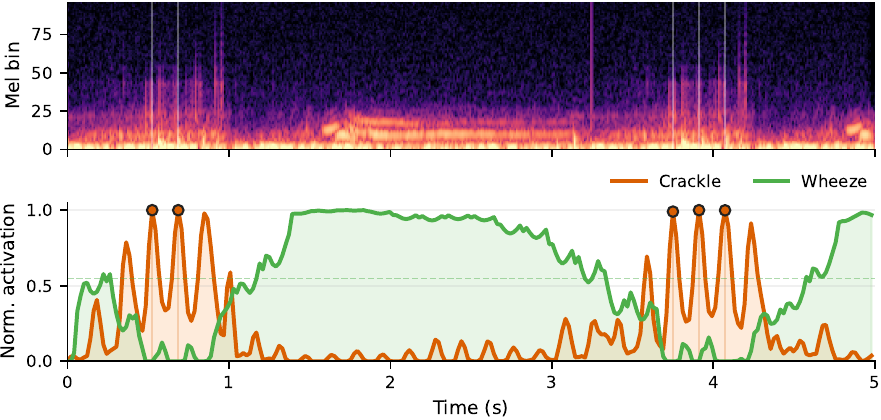}
    \caption{Frame-wise branch responses on a \textit{Both} sample before temporal pooling. The crackle branch produces sparse peak responses around localized transient regions, with selected Top-$K$ frames marked by dots, while the wheeze branch shows broader sustained activations over longer intervals.}
    \label{fig:branch_activation}
\end{figure}

To qualitatively examine whether the learned branches follow the intended acoustic priors, we visualize the frame-wise responses of the branch-specific fused features before temporal pooling. As shown in Fig.~\ref{fig:branch_activation}, a representative \textit{Both} sample containing crackle and wheeze patterns is used. The crackle branch produces sparse peak responses around localized transient regions, and the selected Top-$K$ frames are concentrated on these high-activation segments. In contrast, the wheeze branch exhibits broader and more sustained activations over longer temporal intervals. More importantly, the two responses show clear attribute selectivity: the crackle branch remains relatively inactive during the sustained wheeze-dominant interval, while the wheeze branch does not produce sharp peak responses at most crackle-dominant transient regions. This behavior suggests that the two branches do not simply respond to all abnormal sounds, but instead learn complementary temporal cues aligned with crackle and wheeze attributes. These observations provide qualitative evidence for the effectiveness of the proposed asymmetric temporal aggregation.

\subsection{Backbone Generalization}

\begin{table}[t]
\centering
\footnotesize
\setlength{\tabcolsep}{5pt}
\renewcommand{\arraystretch}{1.08}
\caption{Backbone generalization on ICBHI four-class evaluation.}
\label{tab:backbone_generalization}
\begin{tabular}{l c c c c}
\toprule
Backbone & Method & $S_p$(\%) & $S_e$(\%) & Score(\%) \\
\midrule
\multirow{2}{*}{ResNet-50}
& CE   & \std{76.53}{4.56} & \std{\textbf{32.01}}{2.95} & \std{54.24}{1.00} \\
& AAAF & \std{\textbf{79.11}}{2.02} & \std{31.47}{2.71} & \std{\textbf{55.29}}{0.71} \\
\midrule
\multirow{2}{*}{Eff.-B0}
& CE   & \std{77.63}{3.94} & \std{\textbf{37.72}}{3.19} & \std{57.68}{0.79} \\
& AAAF & \std{\textbf{86.92}}{0.87} & \std{31.05}{1.46} & \std{\textbf{58.98}}{0.66} \\
\midrule
\multirow{2}{*}{AST}
& CE   & \std{73.12}{3.38} & \std{45.91}{2.19} & \std{59.52}{0.84} \\
& AAAF & \std{\textbf{75.27}}{4.46} & \std{\textbf{47.03}}{5.19} & \std{\textbf{61.15}}{0.51} \\
\midrule
\multirow{2}{*}{BEATs}
& CE   & \std{78.77}{3.07} & \std{48.21}{2.32} & \std{63.49}{1.08} \\
& AAAF & \std{\textbf{80.25}}{3.57} & \std{50.35}{3.15} & \std{\textbf{65.30}}{0.50} \\
\bottomrule
\end{tabular}
\end{table}

Table~\ref{tab:backbone_generalization} evaluates the generalization ability of AAAF across different backbone architectures on the ICBHI four-class evaluation. Compared with the standard CE baseline, AAAF consistently improves the overall Score on all evaluated backbones, including CNN-based models, AST, and BEATs. For ResNet-50 and EfficientNet-B0, AAAF increases the Score from 54.24\% to 55.29\% and from 57.68\% to 58.98\%, respectively, mainly by improving Specificity. When applied to AST, AAAF achieves a larger gain of 1.63 percentage points, improving both Specificity and Sensitivity. The best performance is obtained with BEATs, where AAAF improves the Score from 63.49\% to 65.30\%. These results indicate that the proposed task reformulation and acoustic-aware asymmetric modeling are not limited to a specific encoder, while stronger pretrained audio representations further enhance their effectiveness.

\section{Conclusion}
In this paper, we proposed an Acoustic-Aware Asymmetric Framework (AAAF) for respiratory sound classification, which reformulates four-class RSC as two clinically meaningful sound-attribute detection tasks and models crackles and wheezes with branch-specific multi-layer fusion and asymmetric temporal aggregation. Experiments on ICBHI demonstrate strong performance in both four-class classification and two-class abnormality detection, while external validation on SPRSound shows competitive generalization under distribution shifts. Ablation and visual analyses further verify the effectiveness of the task reformulation, acoustic-aware aggregation, and auxiliary supervision. Future work will incorporate clinically relevant metadata and targeted augmentation strategies to further improve robustness and generalization under demographic, acquisition, and class-imbalance variations.

\bibliographystyle{IEEEtran}
\bibliography{mybib}

\end{document}